\pdfoutput=1

\documentclass[11pt]{article}

\usepackage{EMNLP2023}

\usepackage{times}
\usepackage{latexsym}
\usepackage{amssymb}

\usepackage[T1]{fontenc}

\usepackage[utf8]{inputenc}

\usepackage{microtype}
\usepackage{multirow}

\usepackage{inconsolata}
\usepackage[rawfloats=true]{floatrow}
\usepackage{graphicx}

\usepackage{tabularx}

\usepackage{hyperref}
\title{What's happening in your neighborhood? A Multi-lingual Model with Weakly Supervised Approach to Detect Local News}

\author{Deven Santosh Shah\textsuperscript{$^*$},\ Shiying He\textsuperscript{$^*$\textdagger}, \\{\bf Gosuddin Kamaruddin Siddiqi, Radhika Bansal} \\
        Microsoft \\         
        \texttt{\{devenshah,sylviahe,gsiddiqi,rabansal\}@microsoft.com}
        } 

\begin{document}
\maketitle
\begingroup
\renewcommand\thefootnote{\fnsymbol{footnote}}
\footnotetext[1]{These authors contributed equally to this work.}
\footnotetext[2]{Work was done while author was at Microsoft.}
\endgroup

\begin{abstract}
\textit{Local news articles} play a crucial role in informing users about events and issues specific to their geographical regions, ranging from neighborhoods and cities to counties and states. However, accurately detecting whether an article is local news poses a challenge. Naive rule-based methods, such as detecting city names from the news, tend to give erroneous results due to lack of understanding of the news content. In this paper, we present a novel technique empowered by the latest development in natural language processing to generate a training dataset and train a deep learning model, addressing this problem effectively. Our proposed robust framework of data processing encompasses two key components. First, we use weak supervision and domain knowledge for generating and cleaning training data. Second, we leverage large language models like GPT-3 to augment the data for scalable multi-lingual detection. By training a deep learning model with our approach, we achieve high precision and recall in classifying local news, validated on a real-world dataset labeled by human annotators. Our deployed model demonstrates notable enhancements in user engagement as confirmed by online A/B experiments. This contribution advances the field of local news detection in digital news recommendation systems, delivering relevant and timely information to users and fostering local community engagement. 

\end{abstract}

\section{Introduction}
\label{sec:introduction}
Local news plays a significant role in keeping individuals informed about their neighborhoods and fostering community connections. Not only does detecting and showcasing local news to the appropriate audience is crucial for digital news platforms, benefiting both users and publishers and driving user engagement \citep{robindro2017unsupervised} but also it helps support local journalism by getting the local journalist the appropriate audience exposure \citep{gonccalves2021local}. Examples of different types of local news articles include crime reports\footnote{https://www.cbsnews.com/sanfrancisco/news/san-jose-police-arrest-74-year-old-fresno-man-in-connection-to-homicide/}, restaurant openings\footnote{https://sf.eater.com/2022/10/5/23389267/chez-noir-open-new-carmel-restaurant-jonny-black}, real estate trends\footnote{https://www.msn.com/en-us/money/realestate/see-where-home-prices-have-been-rising-the-fastest-in-washington/ss-AA153W3H}, legislative updates\footnote{https://calmatters.org/commentary/2022/10/legislature-must-step-up-for-water-rights-of-all-californians/}, and sports highlights\footnote{https://sports.yahoo.com/samson-ebukam-strong-second-continues-120006770.html}. 

\begin{figure*}[tbh]
\centering
\includegraphics[scale=0.5]{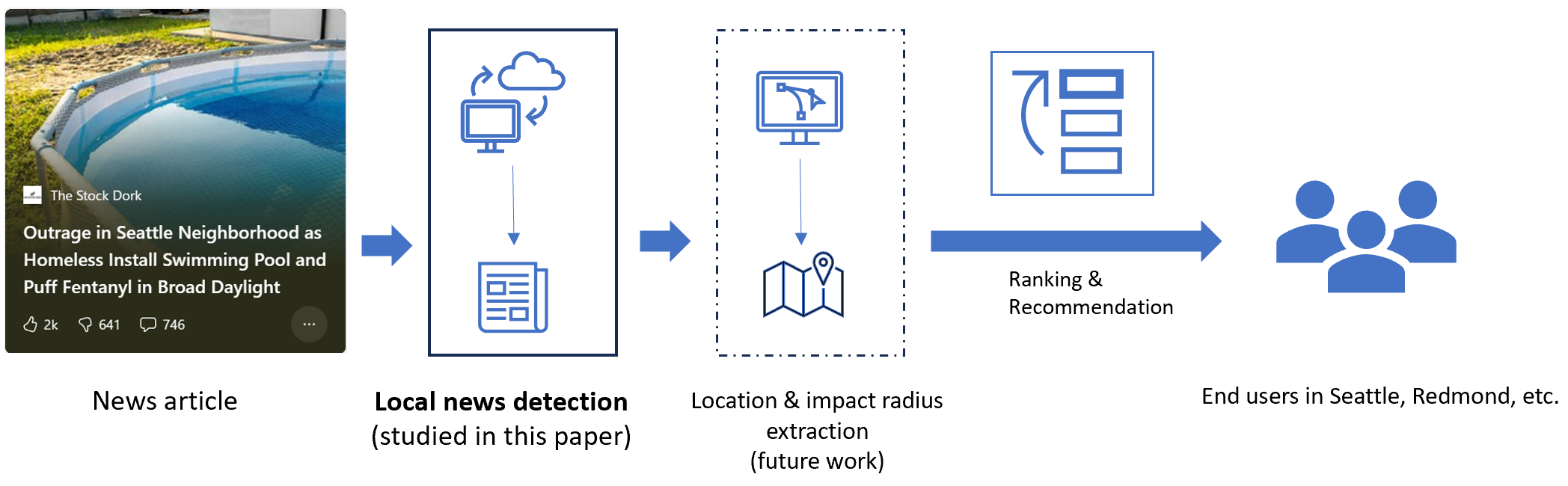} 
\caption{Local News Recommendation}
\label{fig:reco}
\vspace{-10pt}
\end{figure*}

The process of recommending local news involves two steps, highlighted in Figure \ref{fig:reco}: determining if an article\footnote{https://www.thestockdork.com/outrage-in-seattle-neighborhood-as-homeless-install-swimming-pool-and-puff-fentanyl-in-broad-daylight/} is local news and identifying its geolocation and impact radius for targeted delivery to end users. This paper focuses on detecting local news articles that impact users at the neighborhood, city, county, or state level, while the geolocation-specific delivery aspect is left for future research. 

Previous studies explored identifying geolocation information in articles \citep{tahmasebzadeh2021geowine, bell2015system, robindro2017unsupervised, sankaranarayanan2009twitterstand}. However, the presence of geolocation alone does not guarantee the article's local relevance or impact on the community. We discuss the challenges associated with local news detection in section \ref{sec:challenges}. 

\section{Motivation}
\label{sec:challenges}
Detecting local news is a challenging problem. It is not a mere extraction of location names or a certain piece of keywords. It needs a comprehensive summarization of the article, and then predicting the possibility of the news documents that will attract interest from users in particular location affinity. Developing advanced algorithms capable of understanding human language is essential in addressing these challenges. We highlight some drawbacks and challenges associated with relying solely on location names mentioned in articles to determine their local relevance:
\begin{enumerate}
\item \textbf{Articles of national importance:} We encountered news articles where the geolocation was mentioned, but the content was not specific to that location. For example, an article titled “Ryvid Anthem Launch Edition Electric Bike Preorders Are Now Open”\footnote{https://www.rideapart.com/news/604729/ryvid-anthem-launch-preorders-open/} mentions Irvine, CA, but the article's impact extends beyond the local population of Irvine, CA.

\item \textbf{Articles reported from a location: }The geolocation mentioned in the news article does not necessarily indicate that the news is about that location or impact the local community. For example, articles like “Laboratory to study dark matter opens 1km under Australian town”\footnote{https://www.pressreader.com/usa/the-guardian-usa/20220820/282248079355210} and “Prince Harry makes surprise visit to Mozambique ahead of trip back to UK” \footnote{https://metro.co.uk/2022/08/18/prince-harry-makes-surprise-visit-to-mozambique-ahead-of-trip-back-to-uk-17208363/} may mention specific locations but are not specifically about those locations. Such articles often attract a broader readership beyond the localized areas mentioned. Therefore, reporting the location of an article doesn't guarantee it is genuinely local, and showcasing it might not be limited to the local audience.

\item \textbf{Articles with implicitly mentioned location names:} Furthermore, we encountered news articles lacking geolocation information but containing acronyms representing specific locations. For example, articles like “WWU students receive racist emails encouraging violence against Black students”\footnote{https://www.kiro7.com/news/local/video-wwu-students-receive-racist-emails-encouraging-violence-against-black-students/ab9d52e5-75e1-4b13-b8b3-6e0c352ee4d4/} and “SPD updates employee policies on tattoos, jewelry, hair styles, gender language”\footnote{https://komonews.com/news/local/spd-seattle-police-department-employee-policy-tattoo-jewelry-hair-style-beard-gender-language-recruiting-application-officer-king-county},  include acronyms such as WWU (Western Washington University) and SPD (Seattle Police Department). Detecting and mapping these acronyms to the correct locations is a challenging task. Geolocation-based techniques used in previous studies \citep{tahmasebzadeh2021geowine, robindro2017unsupervised, bell2015system, sankaranarayanan2009twitterstand} would miss out on showcasing these hyperlocal news articles to the appropriate audience.
\end{enumerate}

In this paper, we propose a solution to address the problem of local news detection by training a local news classifier trained with a robust data-processing framework. Our contributions include: (1) label correction with weakly supervised data and utilizing knowledge constraints \cite{shah2021distantly} from user click information, and (2) data augmentation for scalability to non-English languages by leveraging GPT-3.

\section{Related Work}
Understanding the geolocation information from texts have attracted research interest in academia and industry. Some literature focused on retrieving location information from news articles and showcasing relevant content to users \citep{tahmasebzadeh2021geowine, bell2015system}. Others explored personalized recommendation based on location data \citep{robindro2017unsupervised, kliman2015location}. Additionally, some papers discussed applications in news generation, such as using Twitter to gather breaking news \citep{sankaranarayanan2009twitterstand} and crowd-sourcing local news through user-generated content \citep{vaataja2012location}. Detailed literature studies can be found in Appendix \ref{sec:appendixcs}.

While geolocation information understanding from news articles has been well studied, as we pointed out in Section \ref{sec:challenges}, the geolocation information alone is not enough to determine whether the news is local or not. To the best of our knowledge, there is no literature on local news classification despite its importance in news recommendation. 





\section{Methodology}
\subsection{Problem Formulation}
In Section \ref{sec:challenges}, we highlighted the need for thorough content analysis to determine the local relevance of news articles. For our purposes, we define \textit{local news} as articles that have an impact on a specific segment of users at the neighborhood, city, county, or state level. To detect local news, we propose a scalable multi-lingual classifier trained on a weakly supervised dataset. In Section \ref{sec:datasetcons}, we will delve into the techniques we utilized to curate this dataset.

\subsection{Model Overview}
\label{sec:modeloverview}
\begin{figure}[tbh!]
\centering
\includegraphics[width=\columnwidth]{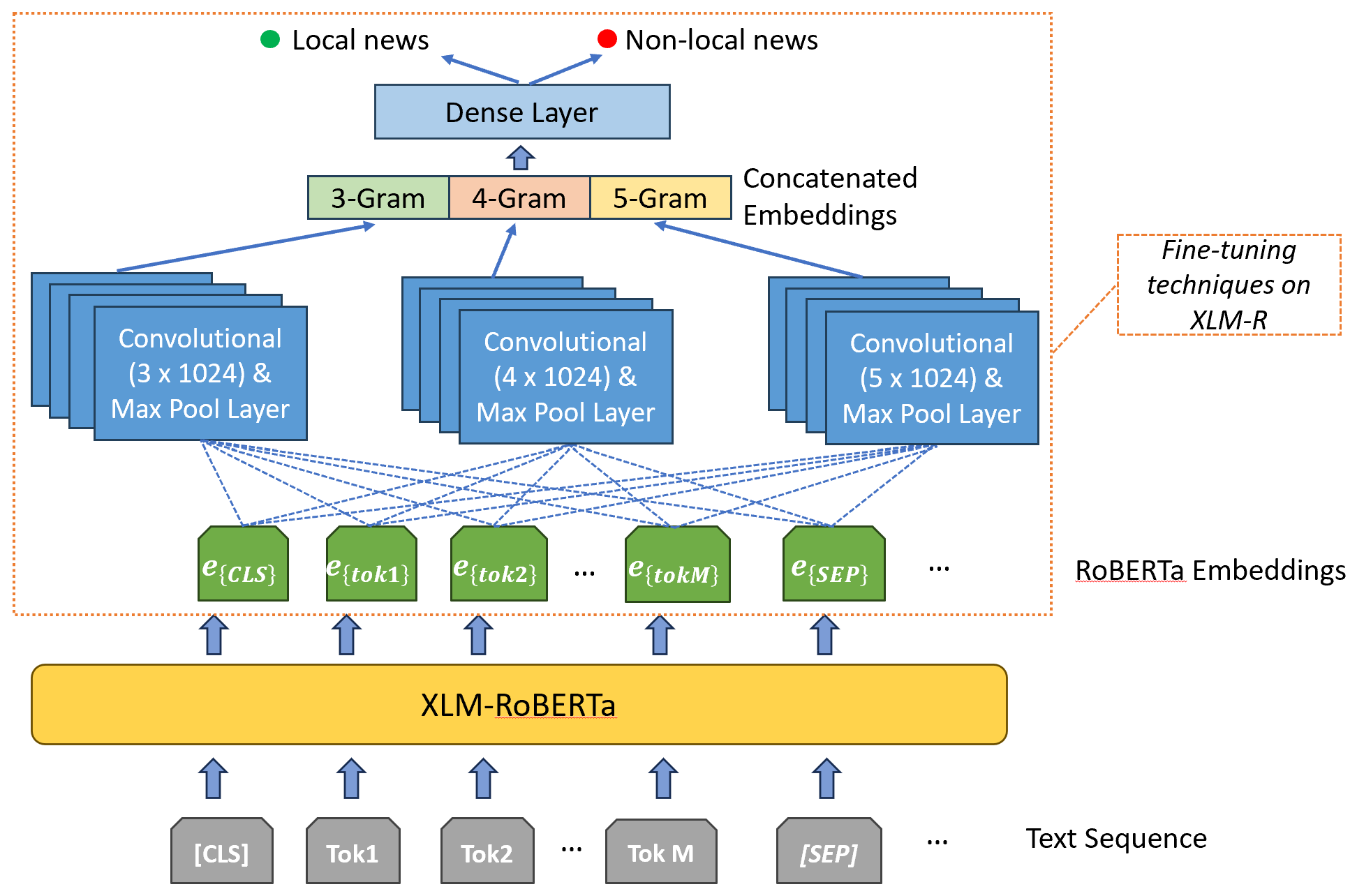}
\caption{Architecture of Multi-lingual Local News Classifier}
\label{fig:dl}
\end{figure}

We first employed XLM-RoBERTa (XLM-R) large-sized model \citep{conneau2019unsupervised} to extract embedding information. XLM-R consists of 24 attention layers, each with 16 attention heads. It was chosen due to its transformer-based architecture and pre-learned associations in 100 different languages, allowing us to transfer the knowledge gained from the English language to other languages efficiently. This multi-lingual capability enables us to scale the classifier across different languages without sacrificing precision and recall on English language population. Then we used convolution layers on top of embeddings from XLM-R to capture 3-, 4-, and 5-grams, which were then concatenated and fed to a dense layer for calculating the probability of an article being a local news piece. We fine-tuned our customized final layers detailed in Section \ref{sec:modeltraining}. The model architecture is shown in Figure \ref{fig:dl}. Further details on the model's scalability to other languages can be found in Section \ref{sec:datasetprepda}.

\subsection{Text Feature Generation}
Four different features were used in the binary classification model for training and inference. 
\begin{itemize}
    \item \textbf{Topics:} We used topics as a representative feature instead of the entire article body. To extract topics, we employed various off-the-shelf and in-house trained models, including keyword extraction, Term Frequency-Inverse Document Frequency (TF-IDF), and  Latent Dirichlet Allocation (LDA).
    \item \textbf{Article Snippet:} We used an in-house trained summarization model to extract the snippet from the body of the article. 
    \item \textbf{Title:} We incorporated the article's title as a feature.
    \item \textbf{URL Features:} URL features improved classifier precision/recall. We split the URL using the ‘/’ delimiter and filtered out domain names and numerical values from URLs. Additionally, if the URL overlapped 80\% with the article title, we removed the title segment.
    
\end{itemize}
The extracted features are combined and separated by the special symbol, $[SEP]$, forming the input for the binary classification model during both training and inference.



\section{Dataset Construction}
\label{sec:datasetcons}
To create a robust training dataset, we applied multiple weakly supervised methods and knowledge constraints \citep{shah2021distantly} for label correction and utilized Neural Machine Translation (NMT) with large language model to augment the positive data.
\subsection{Training label correction}
\label{sec:datasetpreplc}
The raw training labels are obtained from publishers in our in-catalog data, but those labels are very noisy because they are generated based on simple rules on location names and publisher affiliation. Particularly for non-English articles, these labels exhibit significant noise. Therefore, it is necessary to correct the training labels before model training. We implemented the following two techniques for label correction: 
\paragraph{Correction via user click information:}
From our domain knowledge, the articles published by the providers with strong local affinity are mostly local news. Examples of strong local publishers are KIRO Seattle, KING-TV Seattle, etc. We utilized user click logs to assess the affinity of publishers to specific locations and classified them as strong local, strong non-local and ambiguous publishers. Then we re-labeled all the articles from strong local publishers as positive and re-labeled those from strong non-local publishers as negative. To achieve this, we utilized a novel measure called \textbf{gap ratio}. Gap ratio measures the relative \emph{unpopularity} of a publisher’s articles at each city. It is 0 for the city where the publisher’s articles get most clicks and it is 1 for the city where there is no click for the publisher’s articles. The detailed procedure is as follows:

\begin{enumerate}
\item For any publisher $P$, collect all the clicks for articles published from $P$.
\vspace{-5pt}
\item For city $c$ in cities of $\{1,\ldots, C\}$ where the publisher's articles have been impressed, compute the number of clicks from city $c$, denoted as $n_c$.
\item Find $n_{\max} \triangleq \max(n_{1},\ldots, n_{C})$.
\vspace{-5pt}
\item Define \textbf{gap ratio} to be $g_c = 1 - \frac{n_c}{n_{max}}.$
\vspace{-5pt}
\item Compute the number of cities $N$ that has gap ratio smaller than some threshold. In our paper, we use $0.25$, i.e., $N = \left|\{c\in \{1,\ldots, C\}\mid g_c < 0.25\}\right|$. In the high level, $N$ is a robust measure for the number of cities where publisher gets most of its clicks.
\vspace{-5pt}
\item If $N < 5$, we define this publisher as a strong local publisher. If $N > 100$, we define the publisher as a strong non-local publisher. Otherwise, the publisher is defined as ambiguous.
\end{enumerate}
The gap ratio method provides a robust measure to assess the local affinity of different publishers, regardless of their size, whether they are big or small. This knowledge constraint derived from the user click information significantly reduce the noise in the raw labels. 


\paragraph{Correction via transfer learning:} We used a transfer learning process to further reduce noise from the non-English language training dataset. We initially trained a binary local news classifier on English data and obtained its Precision-Recall metrics shown in Table \ref{tab:modelperfpr} in Section \ref{sec:evalpr}. Using GPT-3, we translated non-English articles into English and used the classifer to predict whether it is a piece of local news. If the article's raw label is local but was predicted as strong non-local (local probability $P$ < 0.2) using the English article classifer, we corrected the label to non-local. If the raw label is non-local but was predicted as strong local ($P$ > 0.8), we relabelled it to local.
\subsection{Training data augmentation}
\label{sec:datasetprepda}
\begin{figure}[tbh!]
\vspace{-14pt}
\centering
\includegraphics[scale=0.5]{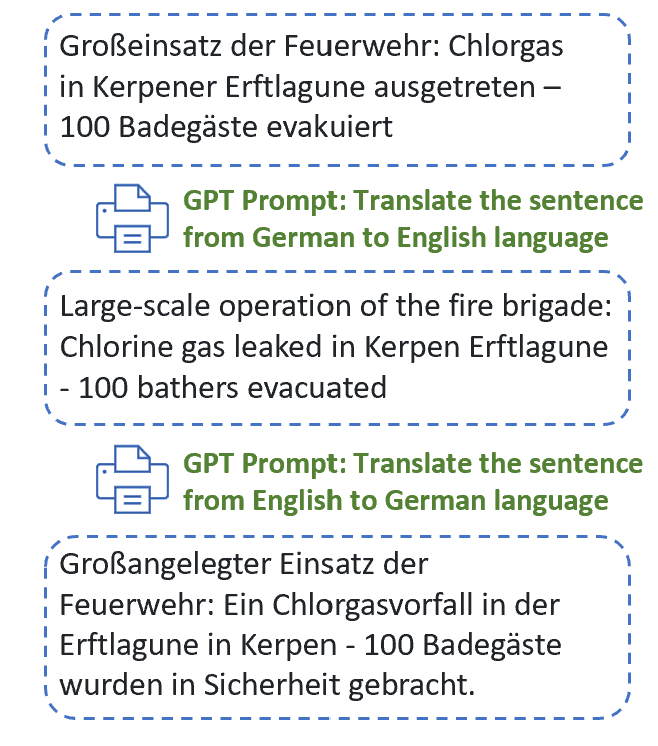} 
\caption{Example of Back-and-Forth Translation}
\label{fig:nmtEg}
\end{figure}

\paragraph{Augmentation via distant supervision: } To augment dataset size and diversity, we integrated more news from outside our data catalog into the data pipeline using a distant supervision method. This was achieved by matching the canonical URL and title of in-catalog articles with those of out-of-catalog articles, aiming to capture similar content and enrich the dataset with diverse features from external sources.
\paragraph{Augmentation via NMT:} We observed that the multi-lingual classifer tends to behave worse on news written in certain languages such as German, Italian, Spanish, Japanese, and French, compared to those written in English. This is due to the scarcity of news articles in those languages. We expanded the non-English news database which enables the model to understand local characteristics in those languages better and prevent label bias during training \citep{shah2019predictive}. Hence We employed GPT-3 for neural machine translation (NMT). Two types of translations are used:
    \begin{itemize}
    \item \textbf{Front Translations:} Local contents were translated from English to the target languages, preserving the precision of the English-trained model.
    \item \textbf{Back Translations:} Data from different languages were translated to English and then back to their original language. This process maintained the semantics of the news articles while introducing variations to expand the model's vocabulary and improve precision and recall on non-English data. An example\footnote{https://www.ksta.de/region/rhein-erft/kerpen/grosseinsatz-der-feuerwehr-chlorgas-in-kerpener-erftlagune-ausgetreten-100-badegaeste-evakuiert-614854} of back-and-forth translations is demonstrated in Figure \ref{fig:nmtEg}.
    \end{itemize}
\subsection{Training data evaluation}
To assess the quality of our proposed data processing methods, we randomly sampled 2673 documents from the generated training data and created a UHRS hitapp\footnote{https://prod.uhrs.playmsn.com/UHRS/} for manual review as ground truth. Table \ref{tab:sampleddistribution} displays the label distributions of the human-reviewed sampled data across different languages and markets. 


\begin{table}[tbh!]
  \begin{center}    
    \scriptsize\begin{tabular}{ccc} 
    \hline
    \textbf{Language} & \textbf{Sample count} & \textbf{Local news \%} \\     
      \hline
      English	& 1556 &	26\%\\
    non-English	& 1117 &	33\%\\ 
    Overall & 2673 & 30\% \\ 
    \hline
      \textbf{Market} & \textbf{Sample count} & \textbf{Local news \%} \\     
      \hline      
        en-AU	& 66 &	19\%\\
        en-CA	& 54 &	39\%\\
        en-GB	& 393 &	33\%\\
        en-IN	& 570 &	29\%\\
        en-US	& 463 & 16\%\\
        de-DE	& 56 & 50\%\\
        es-ES	& 153 &	46\%\\ 
        es-MX	& 157 &	46\%\\       
        fr-FR	& 179 & 41\%\\
        it-IT	& 348 &	47\%\\
        ja-JP	& 224 & 30\%\\
      \hline 
    \end{tabular}
    \caption{Training label distribution from human-reviewed data samples.}
    \label{tab:sampleddistribution}
  \end{center}
  \vspace{-14pt}
\end{table}

We evaluated our proposed method by comparing the precision and recall of the training labels against the true labels determined through human reviews. As shown in Table \ref{tab:reviewedeval}, our data-processing pipeline effectively improved label accuracy. The improvement is especially significant for non-English news, which improved from 50.70\% to \textbf{54.37\%}  for precision and 68.24\% to \textbf{90.59\%} for recall.

\begin{table}[tbh!]
  \begin{center}    
    \scriptsize\begin{tabular}{ccccc} 
      \hline
                                   & \multicolumn{1}{l}{\textbf{Precision}} & \multicolumn{1}{l}{\textbf{Recall}} & \multicolumn{1}{l}{\textbf{Precision}} & \multicolumn{1}{l}{\textbf{Recall}} \\
      \hline
      Overall & 47.13\% & 53.69\% & \textbf{50.35\%} & \textbf{74.84\%}\\
      non-English & 50.70\% & 68.24\% & \textbf{54.37\%} & \textbf{90.59\%}\\
      \hline
    \end{tabular}
    \caption{Evaluation on the generated dataset}
    \label{tab:reviewedeval}
  \end{center}
\vspace{-14pt}
\end{table}

\subsection{Test Dataset}
We conducted manual labeling for the test dataset and utilized the UHRS hitapp for crowdsourcing to obtain labeled data for news articles in both English and non-English languages. The distribution of local news in the test set across different markets is presented in Table \ref{tab:lntestdistribution}.

\begin{table}[tbh!]
  \begin{center}    
    \scriptsize\begin{tabular}{ccc} 
    \hline
      \textbf{Language} & \textbf{Data size} & \textbf{Local news \%} \\
      \hline            
      English & 3446 & 37\% \\
      non-English & 1959 & 46\%\\
      \hline
      \textbf{Market} & \textbf{Data size} & \textbf{Local news \%} \\
      \hline
        en-AU	& 515 &	23\%\\
        en-CA	& 429 &	42\%\\
        en-GB	& 486 &	38\%\\
        en-IN	& 408 &	39\%\\
        en-US	& 1608 & 42\%\\
        de-DE	& 238 &	74\%\\
        es-MX	& 295 &	32\%\\
        es-US	& 268 &	51\%\\
        fr-CA	& 375 &	79\%\\
        fr-FR	& 225 &	47\%\\
        it-IT	& 298 &	84\%\\
        ja-JP	& 260 &	38\%\\

      \hline 
 
    \end{tabular}
    \caption{Test-set distribution and market breakdown.}
    \label{tab:lntestdistribution}
  \end{center}
\end{table}
\vspace{-22pt}
\section{Experiment Setup}
\subsection{Model Training}
\label{sec:modeltraining}
The local news detection model was trained on an Azure compute server equipped with an Intel Xeon Platinum 8168 2.7GHz CPU and 1 Nvidia Tesla V100 32GB GPU. We compiled a dataset of 5.2 million news articles, spanning a period of one month. These articles come from 10 different markets and are available in 6 different languages. 
The distribution of training data across various markets is displayed in Table \ref{tab:lntdistribution}.

\begin{table}[tbh!]
  \begin{center}    
    \scriptsize\begin{tabular}{ccccc} 
      \hline
      \multirow{2}{*}{\textbf{Language}} & \multicolumn{2}{c}{\textbf{w/o NMT}}  & \multicolumn{2}{c}{\textbf{w/ NMT}}                                        \\ 
                                   & \multicolumn{1}{l}{\textbf{Data size}} & \multicolumn{1}{l}{\textbf{local news \%}} & \multicolumn{1}{l}{\textbf{Data size}} & \multicolumn{1}{l}{\textbf{local news \%}} \\
      \hline
      English & 4.5M & 23\% & - & -\\
      non-English & 726K & 49\% & 952K & 60.85\%\\
      \hline
      \textbf{Market} & \textbf{Data size} & \textbf{local news \%} & \textbf{Data size} & \textbf{local news \%}\\
      \hline
        en-AU & 562K & 9\% & - & -\\
        en-CA & 706K & 17\% & - & -\\
        en-GB & 725K & 13\% & - & -\\
        en-IN & 633K & 1\% & - & -\\
        en-US & 1.8M & 41\% & - & -\\
        de-DE & 104K & 27\% & 216K & 64\%\\
        es-MX & 148K & 57\% & 181K & 64\%\\
        fr-FR & 161K & 54\% & 171K & 56\%\\
        it-IT & 211K & 59\% & 213K & 58\%\\
        ja-JP & 100K & 31\% & 168K & 58\%\\  

      \hline 
 
    \end{tabular}
    \caption{Train-set distribution and market-breakdown.}
    \label{tab:lntdistribution}
  \end{center}
  \vspace{-15pt}
\end{table}

In the training process, we kept the pre-trained weights of the XLM-RoBERTa large-sized model \citep{conneau2019unsupervised} and  used cross-entropy as the loss function. We randomly sampled 10 percent data as validation set. Throughout the training process, Adam optimizer was used with batch size 64 and a maximum of 10 epochs. We further used early stopping if the F1-score on the validation set did not improve for 3 epochs.


The whole pipeline is deployed through an online inference service using a single instance equipped with one Nvidia V100 GPU, and two cores of Intel Xeon CPUs, and integrated to our news recommendation platform to serve the users. 

\subsection{Online A/B Experiment}
\label{sec:onlineab}


We conducted online A/B tests to assess the effectiveness of our model on English markets of en-US, en-CA, en-GB, en-AU, and en-IN. Extending the A/B tests to global users will be our future work. In the control group (C), the labeling of local news was done through a rule-based approach, while in the treatment group (T), our proposed model was utilized. In both groups, detecting geo-location and the impact radius of the articles was determined using our in-house rule-based approach using location names and publisher affiliation. The online A/B experiment lasted two weeks. To evaluate the effectiveness of the treatment, we utilize the following metrics: the percentage of local users, local daily active users (DAU), and overall DAU on our news platform. A \textit{local user} is defined as a user who has been shown at least one local document. Our hypothesis is that by detecting the local news more accurately, the recommendation system is able to display more relevant news, thus increasing the percentage of local users as well as DAUs. 


\section{Results and Analysis}
\subsection{Evaluation on Test Dataset}
\label{sec:evalpr}
The model performance on the test set is measured by precision and recall. Our primary focus is on improving precision while maintaining a reasonable recall. Table \ref{tab:modelperfpr} presents precision-recall values for the trained local news classifier models on various datasets, using a prediction score cut-off of 0.5.

The results demonstrate the effectiveness of our proposed data processing methods with GPT-3. The model trained on the generated multi-lingual dataset with the NMT method excels in recall and maintains strong precision on English articles. Moreover, it exhibits promising performance in detecting local news across most market segments.

\begin{table}[tbh!]
  \begin{center}    
    \scriptsize\begin{tabular}{ccccccc} 
      \hline 
      \multirow{2}{*}{\textbf{Language}} & \multicolumn{2}{c}{\textbf{English data}} & \multicolumn{2}{c}{\textbf{Data w/o NMT}}  & \multicolumn{2}{c}{\textbf{Data w/ NMT}} \\ 
      
                                   & \multicolumn{1}{l}{\textbf{P}} & \multicolumn{1}{l}{\textbf{R}} & \multicolumn{1}{l}{\textbf{P}} & \multicolumn{1}{l}{\textbf{R}} & \multicolumn{1}{l}{\textbf{P}} & \multicolumn{1}{l}{\textbf{R}}\\
        \hline
        English & \multicolumn{1}{c}{0.902}   & 0.518     
                         & \multicolumn{1}{c}{0.914}   & 0.472 
                         & \multicolumn{1}{c}{0.883} & \textbf{0.763}   \\                  
        Non-English & \multicolumn{1}{c}{-}   & -    
                        & \multicolumn{1}{c}{0.873}   & 0.298
                        & \multicolumn{1}{c}{\textbf{0.895}}   & \textbf{0.386}   \\
       \hline
       \textbf{Market} & \textbf{P} & \textbf{R} & \textbf{P} & \textbf{R} & \textbf{P} & \textbf{R}\\
      \hline            
      en-US & 0.937 & 0.894 & 0.909 & 0.909 & \textbf{0.952} & \textbf{0.902}\\
      en-CA & 0.918 & 0.647 & 0.904 & 0.541 & 0.864 & \textbf{0.836} \\
      en-GB & 0.843 & 0.654 & 0.881 & 0.485 & \textbf{0.881} & \textbf{0.766} \\
      en-AU & 1.000 & 0.260 & 0.901 & 0.666 & 0.875 & \textbf{0.710} \\
      en-IN & 0.600 & 0.044  & 0.729 & 0.518 & \textbf{0.801} & \textbf{0.720} \\   
      de-DE & - & - & 0.905 & 0.627 & \textbf{0.923}	& 0.438 \\
      es-MX &	- & - & 0.870 & 0.643 & \textbf{0.900}	& 0.739 \\
      es-US &	- & - & 0.824 & 0.628 & 0.820	& \textbf{0.659} \\
      fr-CA &	- & - & 0.862 & 0.788 & \textbf{0.895}	& 0.695 \\
      fr-FR &	- & - & 0.878 & 0.318 & 0.869	& 0.220 \\
      it-IT &	- & - & 0.969 & 0.524 & 0.951	& \textbf{0.530} \\
      ja-JP &	- & - & 0.769 & 0.109 & \textbf{0.846}	& 0.121 \\
     \hline     
    \end{tabular}
    \caption{Model Precision (P) and Recall (R) on test data across languages and breakdown by markets}
    \label{tab:modelperfpr}
  \end{center}
  \vspace{-14pt}
\end{table}
\subsection{A/B testing results}
We report our A/B testing results in Table \ref{tab:abperf}. Recall that those metrics are defined in \ref{sec:onlineab}, in the treatment group (T), where our proposed model was enabled, we observed a significant improvement in the percentage of local users, with an increase of 47.67\%, and an increase of 60.14\% in DAU engaging with local news. Additionally, enabling the new model resulted in a 0.36\% increase in overall DAU on our platform, likely due to the increase in local DAU. All differences are statistically significant with p-values less than 0.05. The results demonstrate that our model leads to higher user engagement with local news content and an overall increase in active users on the platform. 


\begin{table}[tbh!]
  \begin{center}    
    \scriptsize\begin{tabular}{ccc} 
    \hline
    \textbf{Metric} &  \textbf{Delta \%: (T-C)/C} & \textbf{P-Value} \\     
      \hline
      \textbf{Overall DAU}	 &	0.36\% &0.0011\\
      \textbf{Local Users}  & 47.67\% & 0 \\
      \textbf{Local DAU}	 &	60.14\% & 0\\    
    \hline
    \end{tabular}
    \caption{Online A/B test results}
    \label{tab:abperf}
  \end{center}
\vspace{-18pt}
\end{table}
\section{Conclusion}
In this paper, we proposed a method for detecting local news using a multi-lingual model and a robust data-processing framework for training label cleaning and data augmentation with multiple weakly supervised methods. We evaluated our framework on a real-world dataset of news articles in various languages, and demonstrated its effectiveness and scalability in detecting local news with high precision and recall. Moreover, we conducted an online experiment in a digital news recommendation system, and showed that our framework can significantly improve user engagement by providing more relevant and personalized local news content. Our work contributes to the advancement of local news detection and recommendation, and opens up new avenues for future research.



\section*{Limitations}
While our proposed framework, incorporating weakly supervised methods and a large language model, significantly improves the local news detection model in a multi-lingual setting, it does have some limitations that warrant further research. One notable limitation is the scarcity of non-English positive samples in the training data, which may impact the recall of the classification in non-English languages. Another limitation is that the current approach for local news recommendation relies on a rule-based method for detecting the geolocation of the article and it's impact radius. This method may not captue the nuances and variations of geolocation expressions. A possible solution is to use an advanced NLP model for geolocation detection and recognition, which could enhance the accuracy and personalization of local news recommendation. We hope this paper serves as a catalyst for encouraging such advancements in the field of local news and recommendation systems.

\bibliography{local}
\bibliographystyle{acl_natbib}

\appendix

\section{Appendices}
\label{sec:appendix}

\subsection{Literature Studies}
\label{sec:appendixcs}
Several research papers rely on the geolocation information present in the article to detect it as a local news article. \citet{tahmasebzadeh2021geowine} proposed using geolocation and structural type extracted from an image to showcase local news of that area. However, their focus is on news articles belonging to 14 different structural entities like Skyscraper, Square, Historic Site, Waterfall, Tourist Attraction, Museum, Building, Religious Building, Tower, Castle, Bridge, and Monument. The predicted geolocation and the structure type are used to determine the entity from the Wikidata. Hence, their recall decreases to capture only those entities in the Wikidata. They have used ResNets to determine the geolocation in the image, which they have developed as a classification problem making it difficult to scale it worldwide. The entities determined from the Wikidata are used to retrieve news articles from Event Registry. The news articles would get restricted to the structural entities in an area. They won’t be able to capture the local news articles about crime, real estate, science, politics, and weather, to name a few.  

\citet{bell2015system} also presented a technique in which extraction of local news relies only on the geolocation mentioned in the article. Their primary focus was on using Automatic Speech Recognition(ASR) to convert the audio news published by news broadcasting agencies to text and use extractive summarization techniques to extract ten crucial sentences from the text. They also extract the geolocation information using a Named Entity Recognition (NER) model and match it with Open Street Map. However, they do not mention how they would determine if the news article is local. They didn’t mention the process of training the NER model. If they used an off-the-shelf NER \citep{manning2014stanford}, they would miss out on local articles having geolocation information in the form of acronyms like SFPD, WWU, etc., as we mentioned in the Section \ref{sec:introduction}. 

\citet{sankaranarayanan2009twitterstand} proposed using tweets(User Generated Content) on the Twitter platform to gather breaking news in the area. They also proposed an importance score defining a particular news article's importance to a neighborhood. They manually select the users who post the news. They used Naïve Bayes to differentiate it as news or junk. They cluster the tweets and use the geolocation mentioned in the tweet and the user’s location to determine the geolocation foci of the cluster. This foci is considered the geolocation of all the tweets falling in that cluster. The tweet may not have geolocation present; it might not belong to any cluster but will be forced to join a particular cluster and get it geotagged. Or the tweet may belong to a different geolocation than the one it is assigned to in a cluster. It is difficult to scale the framework worldwide as it involves manually selecting the users who post the news. It is difficult to determine the credibility of the users who post on Twitter. This technique is theoretical and not in Production.

\end{document}